# Behavioral Finance Option Pricing Formulas Consistent with Rational Dynamic Asset Pricing

Svetlozar Rachev (Texas Tech University)

Stoyan Stoyanov (Stony Brook University)

Frank J. Fabozzi (EDHEC)

**Summary:** We derive behavioral finance option pricing formulas consistent with the rational dynamic asset pricing theory. In the existing behavioral finance option pricing formulas, the price process of the representative agent is not a semimartingale, which leads to arbitrage opportunities for the option seller. In the literature on behavioral finance option pricing it is allowed the option buyer and seller to have different views on the instantaneous mean return of the underlying price process, which leads to arbitrage opportunities according to Black (1972). We adjust the behavioral finance option pricing formulas to be consistent with the rational dynamic asset pricing theory, by introducing transaction costs on the velocity of trades which offset the gains from the arbitrage trades.

## 1. Introduction

There is a lot of controversy regarding the discrepancy between behavioral finance and rational finance. Rubinstein (2001, p. 16) criticizes the behavioral finance approach to asset pricing. He points out that as trained financial economist, he was taught that the "Prime Directive" in pricing is "Explain asset prices by rational models. Only if all attempts fail, resort to irrational investor behavior." In his opinion, the behavioralists literature "has lost all the constraints of this directive." Statman (1995, p. 14) takes the opposite view, "standard finance is indeed so weighted down with anomalies that it makes much sense to continue the reconstruction of financial theory on behavioral lines." There is an extensive literature that surveys the debate about pros and the cons of behavioral and rational approaches.[1]

In this paper, we reconcile the basic behavioral finance approach to asset pricing with the rational no-arbitrage pricing. We extend the Black (1972) approach to rational dynamic market with two risky assets embedding transaction costs so that the behavioral market model can be viewed as a special case of the Black model. We illustrate our approach using the basic behavioral dynamic asset pricing model proposed by in Shefrin (2005, Section 8.1). The original Shefrin behavioral model allows for arbitrage opportunities when studied from the point of view of the rational finance dynamic asset pricing theory. We demonstrate how to extend Black's approach to encompass the Shefrin behavioral model.

Next, we study the behavioral option pricing model proposed by Benninga and Mayshar (2000). The model assumes that the representative agent applies a discount factor, which is a convex mixture of the discount factors of the two market agents who are sharing an aggregate consumption process. The model allows for arbitrage opportunities. We show that with the introduction of hedging and transaction costs, the Benninga-Mayshar model is arbitrage free within the rational dynamic asset pricing theory.

---

[1] See, among others, Zechhauser (1986), Hirshleifer (2001), Shiller (2003), Barberis and Thaler. (2003, 2005), Brav, Heaton and Rosenberg (2004), Curtis and Statman (2004), Parisi and Smith (2005, Chapter 21), Thaler (2005), and Ricciardi  (2008).

Next, we consider markets with limited arbitrage opportunities, a topic extensively studied in the literature on behavioral finance (see, for example, Chandra (2016, Section 8.3).[2] After introducing the notion of a market agent, who we refer to as the "almost pure arbitrageur", whose Sharpe ratio goes to infinity, we then derive option pricing formula when the almost pure arbitrageur is taking a short position in the option contract. We then introduce hedging costs and consider the limiting case leading to option pricing in the presence of limited arbitrage. All our considerations are based on the rational finance dynamic asset pricing.

In the BF literature the option pricing is relatively new topic[3] with ShM[4] as a cornerstone model built upon the Benninga and Mayshar (2000) model (shortly, BMM). ShM is an equilibrium approach to asset pricing, in which the representative agent views the return from the underlying asset as a mixture of two different normal distributed returns representing the heterogeneous views on the asset return of the buyer and the seller of the option. Since mixture of two different log-normal distributions is not infinitely divisible[5] , the price process of the representative investor is not a semimartingale and thus the model allows for arbitrage opportunities[6].

---

[2] Grossman and Stiglitz (1980) argue that in a perfectly efficient market (a market with no arbitrage opportunities), traders would not have the incentive to gather information, and all gathered information would have been costless.

[3] Shefrin (2005), Chapter 21, Locke (2010), Pena, Alemanni and Zanotti (2011), Matsumura K. and Kawamoto M.(2013).

[4] See Shefrin (2005) Chapters 8 and 21

First let us recall that lognormal distributions are infinitely divisible, but if the mixing measure is with finite support and thus the mixing measure is not infinitely divisible, the mixture of log normal distributions is not infinitely divisible distribution. This, unfortunately, leads to the fact that the process of the representative agent's price process as defined in ShM (formula (8.15) on page 103, and formula (21.7) on page 306) is not infinitely divisible process in the limit (when the number of steps to the terminal time increase to infinity). As a result, ShM is not free of arbitrage opportunities Similar problem arises in BMM where it is written" The function $f^*(Y)$ in (41) can be considered as an average of the two agents' density functions." Indeed, however $f^*(Y)$ is not infinitely divisible density and that will lead to arbitrage opportunities BMM. Neither ShM nor BMM provide hedging strategies as those cannot be constructed.

Second, within the Rational Dynamic Asset Pricing Theory (RDAPT) the most important problem is the characterization of economically rational consistent models for financial markets[7]. In the

[5]  For extensive reviews on infinitely divisible distributions and processes we refer to Bondesson (1992), Sato (1999), Bose, Dasguota and Rubin (2002), Steutel and van Harn (2004)., Kyprianou (2006), Applebaum (2009).

[6]  Shefrin (2005) page 319, formulas (21.24), (21.25) defines a market model with two investors sharing two price processes with common Brownian motion as market driver, the same volatility parameters and different instantaneous mean returns, which leads to arbitrage opportunities according to Black (1972).

[7] See for example Duffie (2001), Chapter 6.

RDAPT, the central assumption is that of no-arbitrage: a market participant (designated as ⊐) should not engage in a contract in which ⊐ can lose infinite amount of money in a frictionless market. Regardless how irrational ⊐ could be, ⊐ should not be so misled as to be subject to infinite loss. There is no reason for behavioralists to object that no-arbitrage assumption as a fundamental notion in finance, rational or behavioral. If this assumption is not satisfied, agents using financial asset pricing formulas allowing for arbitrages could suffer tremendous losses. If a trader being a behaviorist decides still to apply the ShM-or BMM-  option pricing while being long in the contract, there will be a "rational" trader who will take the short position and apply arbitrage strategy. Indeed, in ShM-  and BMM--approaches there is no suggestion for a hedging strategy of the option seller should use, simply because there is none.

The failing in the ShM-option pricing, viewed from RDAPT viewpoint, is due to the fact that the price process of the representative investor is not a semimartingale. The main reasons to use semimartingales in modeling the dynamics of the asset prices are the following [8]:

(1) The semimartingales are the largest possible class for price processes, when defining the gains from the trading strategies applied to price processes[9];

---

[8] Black and Scholes (1973), Merton (1973), Harrison and Kreps (1979), Harison and Pliska (1981), (1983), Dalang, Morton and Willinger (1990), Ansel and Stricker,(1994), Delbaen and Schachermayer (1994),(1997), (1998), (2006), Rachev et al. (2011), Strong (2014). .

[9] We do not discuss fractional processes (see Mishura (2008) and other generalizations (see for example, Frittelli (1997) and Kardaras (2010))   as they are not related to the problem we are dealing with.

(2)A second reason why semimartingale models are omnipresent is the fundamental work in RDAPT on no-arbitrage criteria which can be summarized as Fundamental Asset Pricing Theorem and "No Free Lunch with Vanishing Risk" (NFLVR) condition — and the mathematical notion of existence of equivalent probability measures, under which asset prices have some sort of martingale property, which leads to the price process again has to be a semimartingale.

In this paper, we suggest several approaches to adjust the ShM- BMM- option pricing formulas for traders having heterogeneous views on the underlying pricing process so that those formulas are co0nsisting with the RDAPT. Namely, we will impose trading costs (so called **arb-costs**) which will offset the gains from the arbitrage opportunities the hedger could have. Equilibrium options pricing formulas when the traders have heterogeneous beliefs are well studied in RDART[10]. Our approach to option pricing in the presence of heterogeneous beliefs is different, which can be roughly explained as follows: for the hedger ⊐ to realize an arbitrage strategy, ⊐ must trade in high speed, thus placing arb-costs on the velocity [11] of trades can offset the gains ⊐ accumulates when applying the arbitrage trade. In regular hedging when no arbitrage occurs the arb-costs are not significant. As a conclusion, our approach leads to the opinion that in modern financial markets, transaction cost on the velocity of trading should be imposed to remove potential arbitrage gains.

---

[10] Buraschi and Jiltsov (2p006), Chabakauri G. (2013), He and Shi (2016). Muhle-Karbe and Nutz (2016) showed simple wining strategies in option contracts when the buyer and the seller of the option contract have heterogeneous views.

[11] See Duffie (2001), p 104, formula (3).

The paper is organized as follows. In Section 2 we adjust ShM to be consistent with RDAPT applying binomial tree model with transaction costs. In Section 3 introduce arb-costs on the delta- hedge positions in order to eliminate the gains from the arbitrage opportunities, ShM model can invoke in continuous time asset pricing. In Section 4 we consider similar type arb- costs on the trading velocity using binomial tree model. In Section 5 we consider a general model with arb- costs on the delta- and gamma-positions of the hedge portfolio. Our concluding remarks are in Section 6.

## 2. Shefrin's behavioral asset pricing model with transaction costs

We start with the description of ShM but from the viewpoint of RDAPT. Consider a financial market with two investors sharing an aggregate consumption (AC) $\omega(0) > 0$ amount at $t^{(0)}$. At any subsequent period $t^{(k+1)} = (k+1)\Delta t, k = 0,1,..,n-1, t^{(n)} = T, n \uparrow \infty$, the aggregate amount available will unfold through a binomial process, growing by $u^{(\Delta t)} > 1$, or $d^{(\Delta t)} < 1$. Under ShM the investor $\beth^{(j)}, j = 1,2$ attaches probability $p^{(j)}(\Delta t)$ for upward movement of the AC-process and $1 - p^{(j)}(\Delta t)$ for downturn movement, that is, the discrete AC-dynamics is given by

$$\omega^{(j)}\big(t^{(k+1)}\big) = \begin{cases} \omega^{(j)}\big(t^{(k)}\big)u^{(\Delta t)}, w.p. \, p^{(j)}(\Delta t) \\ \omega^{(j)}\big(t^{(k)}\big)d^{(\Delta t)}, w.p. \, 1 - p^{(j)}(\Delta t). \end{cases} \tag{1}$$

The ShM is an equilibrium model based on the heterogeneous beliefs of the two investors or determine the dynamics of the state-price process of the representative investor. The ShM derives the dynamics of the state-price process of the representative agent as a probability mixture of different geometric Brownian motions representing the dynamics of the state-price processes of two investors. Thus, the state-price process of the representative agent is not a semimartingale and ShM allows for arbitrage opportunities.

## 2.1. *Binomial tree with heterogeneous views on the model parameters*

Our goal, is to adapt the behavioral framework in the ShM, within RDAPT. First, let us make the following observation when extending the ShM. Suppose $\beth^{(j)}, j = 1,2,$ differ not only on their views of the probabilities $p^{(j)}(\Delta t)$ but also on their views of the size of the upward and downward movements, that is, $\beth^{(j)}$ views the AC-process as follows,

$$\omega^{(j)}\big(t^{(k+1)}\big) = \begin{cases} \omega^{(j)}\big(t^{(k)}\big)u^{(j,\Delta t)}, w.p.\, p^{(j)}(\Delta t) \\ \omega^{(j)}\big(t^{(k)}\big)d^{(j,\Delta t)}, w.p.\, 1 - p^{(j)}(\Delta t), j = 1,2 \end{cases} \qquad (2)$$

Let us now make use of Kim at al (2016)'s extension of the CRR[12] binomial tree, assuming that

$$\begin{cases} p^{(j)}(\Delta t) := g^{(j)} + v^{(j)}\sqrt{\Delta t}, g^{(j)} \in (0,1), v^{(j)} \in R \\ u^{(j,\Delta t)} := 1 + \gamma^{(j)}\Delta t + \sqrt{\frac{1-p^{(j)}(\Delta t)}{p^{(j)}(\Delta t)}}\sigma^{(j)}\sqrt{\Delta t}, \gamma^{(j)} \in R, \sigma^{(1)} > \sigma^{(2)} > 0 \\ d^{(j,\Delta t)} := 1 + \delta^{(j)}\Delta t - \sqrt{\frac{p^{(j)}(\Delta t)}{1-p^{(j)}(\Delta t)}}\sigma^{(j)}\sqrt{\Delta t}, \delta^{(j)} \in R, j = 1,2 \end{cases} \qquad (3)$$

Then the bivariate binomial tree (2) generates a discrete pricing process in Skorokhod space $D[0,T]^2$, which converges (as $\Delta t \downarrow 0, n \uparrow \infty, n\Delta t = T$) to a bivariate geometric Brownian motion

$$\omega^{(j)}(t) = \omega(0)\exp\left(\left(\mu^{(j)} - \frac{\sigma^{(j)^2}}{2}\right)t + \sigma^{(j)}B(t)\right), t \geq 0, \qquad (4)$$

where $\mu^{(j)} = g^{(j)}\gamma^{(j)} + \big(1 - g^{(j)}\big)\delta^{(j)}$ and $B(t), t \geq 0$ is a Brownian motion generating a stochastic basis $(\Omega, \mathcal{F}, \mathbb{F} = (\mathcal{F}_t, t \geq 0), \mathbb{P})$. . Furthermore, the risk-neutral dynamics $\omega^{(\mathbb{Q},j)}(t), t \geq 0$, will be determined by

$$\omega^{(\mathbb{Q},j)}(t) = \omega(0) \exp\left(\left(r - \frac{\sigma^{(j)^2}}{2}\right)t + \sigma^{(j)} B^{\mathbb{Q}}(t)\right), t \geq 0,, j = 1,2, \tag{5}$$

where $B^{\mathbb{Q}}(t), t \geq 0$ is a Brownian motion generating a stochastic basis $(\Omega, \mathcal{F}, \mathbb{F} = (\mathcal{F}_t, t \geq 0), \mathbb{Q})$, $dB^{\mathbb{Q}}(t) = dB(t) + \frac{\mu^{(j)} - r}{\sigma^{(j)}} dt$ on $\mathbb{P}$. Per Black (1972) model, the riskless rate $r$, is given by

$$r = \frac{\mu^{(1)}\sigma^{(2)} - \mu^{(2)}\sigma^{(1)}}{\sigma^{(2)} - \sigma^{(1)}}. \tag{6}$$

---

[12] Cox,Ross and Rubinstein (1979).

Then following the ShM, suppose that at $t = 0$, the aggregate wealth of the $jth$ investor is $W^{(j,0)}, j = 1,2, W^{(1,0)} + W^{(1,0)} = W^{(0)}$, and the relative wealth at $t = 0$, is $\omega^{(j)}(0) = \frac{W^{(j,0)}}{W^{(0)}}$, $\omega^{(1)}(0) + \omega^{(2)}(0) = 1$.

The risk-neutral approach (5), (6) provides an additional characterization of ShM under the assumption (2), (3): Suppose that ℷ$^{(j)}$, 1,2, share the cumulative return of the AC-process in proportions $\alpha = (\alpha^{(1)}, \alpha^{(2)})$, $\alpha^{(j)} \in R \setminus \{0\}, \alpha^{(1)} + \alpha^{(2)} = 1$, that is, the combined cumulative return of the both investors is:

$$\alpha^{(1)} ln\omega^{(1)}(t) + \alpha^{(2)} ln\omega^{(2)}(t) = lnG^{(\alpha)}(t), t \geq 0. \tag{7}$$

Hence, $\alpha$ is an acceptable (arbitrage-free) allocation if and only if $\mathfrak{G}^{(\alpha)}$ is a perpetual derivative with price process:

$$G^{(\alpha)}(t) = g\left(\omega^{(1)}(t), \omega^{(2)}(t)\right), t \geq 0, g(x_1, x_2) = x_1^{\alpha^{(1)}} x_2^{\alpha^{(2)}}, x_j > 0, j = 1,2, \tag{8}$$

To find the acceptable value for $\alpha$, notice first that according (5) and (6), $g(x_1, x_2) = x_1^{\alpha^{(1)}} x_2^{\alpha^{(2)}}, x_j > 0, j = 1,2$ satisfies the partial differential equation (PDE:

$$\frac{\partial g(x_1, x_2)}{\partial x_1} x_1 + r \frac{\partial g(x_1, x_2)}{\partial x_2} x_2 - r g(x_1, x_2) +$$

$$+ \frac{1}{2} \frac{\partial^2 g(x_1, x_2)}{\partial x_1^2} x_1^2 \sigma^{(1)2} + \frac{1}{2} \frac{\partial^2 g(x_1, x_2)}{\partial x_2^2} x_2^2 \sigma^{(2)2} + \frac{\partial^2 g(x_1, x_2)}{\partial x_1 \partial x_2} x_1 x_2(t) \sigma^{(1)} \sigma^{(2)} = 0. \tag{9}$$

Thus, $\alpha = \left( \alpha^{(1)}, \alpha^{(2)} \right)$, should satisfy:

$$r\left( \alpha^{(1)} + \alpha^{(2)} - 1 \right) + \frac{1}{2} \alpha^{(1)} \left( \alpha^{(1)} - 1 \right) \sigma^{(1)2} + \frac{1}{2} \alpha^{(2)} \left( \alpha^{(2)} - 1 \right) \sigma^{(2)2} +$$

$$+ \alpha^{(1)} \alpha^{(2)} \sigma^{(1)} \sigma^{(2)} = 0. \tag{10}$$

Because $\alpha^{(j)} \in R \setminus \{0\}, \alpha^{(1)} + \alpha^{(2)} = 1$, then $\sigma^{(1)} = \sigma^{(2)} = \sigma$, with riskless rate $r$ in (6) exploding to $\pm\infty$ as the derivative $\mathfrak{G}^{(\alpha)}$ is an arbitrage security. This shows that ShM leads to pricing model with arbitrage opportunities.

To illustrate the problem with the ShM again, let us return to the binomial model (1), when $u^{(j, \Delta t)} = u^{(\Delta t)} > 1 > d^{(j, \Delta t)} = d^{(\Delta t)}$, $\Delta t \downarrow 0, n \uparrow \infty, n\Delta t = T$. We require[13] that the first two moments of the $\Delta t$- increments of processes (1) and (2) coincide, that is $u^{(\Delta t)} > 1$, or $d^{(\Delta t)} < 1$.

$$u^{(\Delta t)} p^{(j)}(\Delta t) + d^{(\Delta t)} \left( 1 - p^{(j)}(\Delta t) \right) =$$

$$= \mathbb{E} \exp\left( \left( \mu^{(j)} - \frac{\sigma^{(j)2}}{2} \right) \Delta t + \sigma^{(j)} B(\Delta t) \right) = 1 + \mu^{(j)} \Delta t, \tag{11}$$

and

---

[13] All terms of order $o(\Delta t)$ are omitted.

$$u^{(\Delta t)^2} p^{(j)}(\Delta t) + d^{(\Delta t)^2}\left(1 - p^{(j)}(\Delta t)\right) =$$

$$= \mathbb{E}\exp\left(2\left(\mu^{(j)} - \frac{\sigma^{(j)^2}}{2}\right)\Delta t + 2\sigma^{(j)}B(\Delta t)\right) = 1 + \left(2\mu^{(j)} + \sigma^{(j)^2}\right)\Delta t. \tag{12}$$

We search for solution of (11) and (12) in the general type: $u^{(\Delta t)} = 1 + a\Delta t + \mathscr{b}\sqrt{\Delta t}$, $d^{(\Delta t)} = 1 + c\Delta t - d\sqrt{\Delta t}$, $p^{(j)}(\Delta t) = p^{(j)} + \hbar^{(j)}\sqrt{\Delta t} + \mathscr{g}^{(j)}\Delta t$. Then, from (8) and (9), it can be shown that $p^{(j)} = \frac{1}{2}$, $\mathscr{b} = d = \sigma^{(j)}$, $h^{(j)} = \frac{\mu^{(j)}}{b+d} - a\frac{d}{(b+d)^2} + c\frac{d}{(b+d)^2}$. Furthermore, as $\Delta t \downarrow 0$, $n \uparrow \infty$, $n\Delta t = T$, the dynamics (1) weakly converges to the dynamics of bivariate GBM:

$$\omega^{(j)}(t) = \omega(0)\exp\left(\left(\mu^{(j)} - \frac{\sigma^2}{2}\right)t + \sigma^2 B(t)\right), t \geq 0, j = 1,2. \tag{13}$$

### 2.2. _Binomial tree with transaction costs eliminating the gains from arbitrage trades_

Unfortunately, the dynamics (13) allows for arbitrages opportunities as soon as $\mu^{(1)} \neq \mu^{(2)}$. To reconcile this major discrepancy between the behavioral asset pricing and RDAPT we introduce trading cost to eliminate the gain from potential arbitrage opportunities. We define the following AS-process on a tree with transaction cost[14]: as $\Delta t \downarrow 0$,

---

[14] Having all terms of order $o(\Delta t)$ are omitted, and if $\mathbb{c}^{(j)} = 1$ and $\varepsilon = 0$, then (14) is the CRR-binomial tree model (see Cox, Ross, Rubinstein (1979))

$$\omega^{(*,j)}\big(t^{(k+1)}\big) =$$

$$\omega^{(*,j)}\big(t^{(k)}\big)\begin{cases}\left\{1 + \mathbb{c}^{(j)}\sigma\sqrt{\Delta t} + \frac{1}{2}\big(\mathbb{c}^{(j)}\sigma\big)^2\Delta t\right\}, w.p.\ \varepsilon^{(j)}\ p^{(j)}(\Delta t)\\\left\{1 + \sigma\sqrt{\Delta t} + \frac{1}{2}\sigma^2\Delta t\right\}, w.p.\ \big(1 - \varepsilon^{(j)}\big)\ p^{(j)}(\Delta t)\\\left\{1 - \sigma\sqrt{\Delta t} + \frac{1}{2}\sigma^2\Delta t\right\}, w.p.\ \big(1 - \varepsilon^{(j)}\big)\big(1 - p^{(j)}(\Delta t)\big)\\\left\{1 - \mathbb{c}^{(j)}\sigma\sqrt{\Delta t} + \frac{1}{2}\big(\mathbb{c}^{(j)}\sigma\big)^2\Delta t\right\}, w.p.\ \varepsilon^{(j)}\big(1 - p^{(j)}(\Delta t)\big), j = 1,2,\end{cases}\tag{14}$$

where

$(i)$  $p^{(j)}(\Delta t) = \frac{1}{2} + \frac{\mu^{(j)} - \frac{\sigma^2}{2}}{2\sigma}\sqrt{\Delta t}$;

$(ii)$  $\mathbb{c}^{(j)} > 1$,  is the transaction rate available to investor $\beth^{(j)}, j = 1,2$ (we assume that $\beth^{(j)}, j = 1,2$ have heterogeneous transaction rates:  $\mathbb{c}^{(1)} \neq \mathbb{c}^{(2)}$);

$(iii)$  $\varepsilon^{(j)} \in (0,1), \varepsilon^{(1)} \neq \varepsilon^{(2)}$.

To determine the continuous-time dynamics $\omega^{(*,j)}(t), t \geq 0$, derived from (14) as $\Delta t \downarrow 0$,

$$\omega^{(*,j)}(t) = \omega(0)\exp\left(\left(m^{(j)} - \frac{v^{(j)2}}{2}\right)t + v^{(j)}B(t)\right), t \geq 0, m^{(j)} \in R, v^{(j)} > 0, j = 12,\tag{15}$$

let us match the first two moments of  $\omega^{(j)}(\Delta t)$ and $\omega^{(*,j)}(\Delta t)$. We readily obtain

$$m^{(j)} = \mu^{(j)} + \frac{1}{2}\sigma^2 c^{(j)}\big(c^{(j)} - 1\big)\varepsilon^{(j)},\ \ v^{(j)2} = \sigma^2\big(1 + \big(c^{(j)} - 1\big)\varepsilon^{(j)}\big), j = 1,2,\tag{16}$$

and thus,

$$\omega^{(*,j)}(t) = \omega(0)\exp\left(\left(\mu^{(j)} - \frac{1}{2}\sigma^2\right)t + \sigma\sqrt{1 + \big(c^{(j)} - 1\big)\varepsilon^{(j)}}B(t)\right), t \geq 0, j = 1,2.\tag{17}$$

Furthermore, bivariate binomial tree (14) generates a discrete pricing process in Skorokhod space $D[0,T]^2$, which converges (as  $\Delta t \downarrow 0, n \uparrow \infty, n\Delta t = T$) to a bivariate geometric Brownian motion (17).

Now, following ShM (1), investor $\beth^{(j)}, j = 1,2$, views AC-process $\omega^{(j)}(t), t \geq 0$ as determined by(13). However, $\beth^{(j)}$'s trades are subject to transaction costs and as a result, $\beth^{(j)}$ trades the AC-process as $\omega^{(*,j)}(t), t \geq 0$ determined by (17). Next, consider again the perpetual derivative, $\mathfrak{G}^{(\alpha)}$ is with price process, $G^{(\alpha)}(t), t \geq 0$, given by (8). Then the $G^{(\alpha)}(t) = g\left(\omega^{(1)}(t), \omega^{(2)}(t)\right), t \geq 0$ dynamics is given by the Itô formula

$$dG^{(\alpha)}(t) = dg\left(\omega^{(1)}(t), \omega^{(2)}(t)\right) =$$

$$= \left\{ \begin{array}{c} \frac{\partial g\left(\omega^{(1)}(t), \omega^{(2)}(t)\right)}{\partial x_1} \omega^{(1)}(t)\mu^{(1)} + \frac{\partial g\left(\omega^{(1)}(t), \omega^{(2)}(t)\right)}{\partial x_2} \omega^{(2)}(t)\mu^{(2)} + \\ \frac{1}{2}\frac{\partial^2 g\left(\omega^{(1)}(t), \omega^{(2)}(t)\right)}{\partial x_1^2}\left(\omega^{(1)}(t)\right)^2 \sigma^2 + \frac{1}{2}\frac{\partial g^2\left(\omega^{(1)}(t), \omega^{(2)}(t)\right)}{\partial x_2^2}\left(\omega^{(2)}(t)\right)^2 \sigma^2 + \\ \frac{\partial^2 g\left(\omega^{(1)}(t), \omega^{(2)}(t)\right)}{\partial x_1 \partial x_2} \omega^{(1)}(t)\omega^{(2)}(t)\sigma^2 \end{array} \right\} dt +$$

$$+ \left\{ \frac{\partial g\left(\omega^{(1)}(t), \omega^{(2)}(t)\right)}{\partial x_1} \omega^{(1)}(t) + \frac{\partial g\left(\omega^{(1)}(t), \omega^{(2)}(t)\right)}{\partial x_2} \omega^{(2)}(t) \right\} \sigma B(t). \qquad (18)$$

Consider a self-financing strategy $A^{(j)}(t), t \geq 0, j = 1,2$,

$$g\left(\omega^{(1)}(t), \omega^{(2)}(t)\right) = A^{(1)}(t)\omega^{(1)}(t) + A^{(2)}(t)\omega^{(2)}(t). \qquad (19)$$

Due to the transaction costs the strategy $A^{(j)}(t), t \geq 0, j = 1,2$ detremines the following dynamics of the replicating portfolio:

$$dg\left(\omega^{(1)}(t), \omega^{(2)}(t)\right) = \{A^{(1)}(t)\omega^{(1)}(t)\mu^{(1)} + A^{(2)}(t)\omega^{(2)}(t)\mu^{(2)}\}dt +$$

$$+\{A^{(1)}(t)\omega^{(1)}(t)v^{(1)} + A^{(2)}(t)v^{(2)}(t)\sigma^{(2)}\}dB(t). \qquad (20)$$

Equating the terms with $dg\left(\omega^{(1)}(t), \omega^{(2)}(t)\right)$, leads to:

$$\begin{cases} A^{(1)}(t)\omega^{(1)}(t) = \dfrac{\frac{\partial g\left(\omega^{(1)}(t),\omega^{(2)}(t)\right)}{\partial x_1}\omega^{(1)}(t)\sigma + \frac{\partial g\left(\omega^{(1)}(t),\omega^{(2)}(t)\right)}{\partial x_2}\omega^{(2)}(t)\sigma - g\left(\omega^{(1)}(t),\omega^{(2)}(t)\right)v^{(2)}}{v^{(1)}-v^{(2)}} \\[3mm] A^{(2)}(t)\omega^{(2)}(t) = \dfrac{g\left(\omega^{(1)}(t),\omega^{(2)}(t)\right)v^{(1)} - \frac{\partial g\left(\omega^{(1)}(t),\omega^{(2)}(t)\right)}{\partial x_1}\omega^{(1)}(t)\sigma - \frac{\partial g\left(\omega^{(1)}(t),\omega^{(2)}(t)\right)}{\partial x_2}\omega^{(2)}(t)\sigma}{v^{(1)}-v^{(2)}} \end{cases} \quad (21)$$

Combining (18) - (21) results in the following PDE for $g(x_1, x_2), x_1 > 0, x_2 > 0$ :

$$\left(r^{(*)} + C_y^{(1)}\right)\frac{\partial g(x_1,x_2)}{\partial x_1}x_1 + \left(r^{(*)} + C_y^{(2)}\right)\frac{\partial g(x_1,x_2)}{\partial x_2}x_2 - r^{(*)}g(x_1,x_2) +$$

$$+\frac{1}{2}\frac{\partial^2 g(x_1,x_2)}{\partial x_1^2}x_1^2\sigma^2 + \frac{1}{2}\frac{\partial^2 g(x_1,x_2)}{\partial x_2^2}x_2^2\sigma^2 + \frac{\partial^2 g(x_1,x_2)}{\partial x_1 \partial x_2}x_1 x_2(t)\sigma^2 = 0, \quad (22)$$

where

$$r^{(*)} = \frac{\mu^{(1)}v^{(2)} - \mu^{(2)}v^{(1)}}{v^{(2)} - v^{(1)}}, C_y^{(j)} = v^{(j)}\frac{\mu^{(1)} - \mu^{(2)}}{v^{(1)} - v^{(2)}} - \sigma\frac{m^{(1)} - m^{(2)}}{v^{(1)} - v^{(2)}}, j = 1,2. \quad (23)$$

Thus, arbitrage-free wealth allocation (7) in the presence of transaction costs, is given by $\alpha^{(1)} + \alpha^{(2)} = 1$, where

$$\alpha^{(1)}\alpha^{(2)}\sigma^2 - C_y^{(1)}\alpha^{(1)} - C_y^{(2)}\alpha^{(2)} = 0. \quad (24)$$

We summarize our findings in the following proposition.

*PROPOSITION 1: Consider a financial market with two investors sharing an aggregate consumption (AC) amount $\omega(0) = 1$ at $t = 0$. Two investors $\beth^{(j)}, j = 1,2 = 0$, share initial wealth $\omega(0)$ with allocation weight $\alpha^{(j)} \in R, j = 1,2$, respectively, $\alpha^{(1)} + \alpha^{(2)} = 1$. Investor $\beth^{(j)}$ views the dynamics of the AC-process as (13), but due to transaction cost, $\beth^{(j)}$ trades the AC-process under the dynamics (17). Then the no-arbitrage allocation $(\alpha^{(1)}, \alpha^{(2)})$ is given by (24). The no-arbitrage riskless rate $r^{(*)}$ and $\beth^{(j)}$- transaction yield $C_y^{(i)}$ are given by (23).*

Proposition 1 bridges the behavioral asset pricing approach in ShM with the RDAPT in the presence of transaction costs.

Shefrin (2005) discusses various behavioral phenomena leading to asymmetric non-Gaussian return and volatility clustering as well as momentum (long-range dependence). Those phenomena can be encompassed in our general setting, extending the dynamics defined (15) and (17) with Barndorff-Nielsen -Ornstein-Uhlenbeck-type wealth process dynamics[15].

3. **Option Pricing with Heterogeneous Views on the Underlying Asset in the Presence of Arb-transaction Costs**

We start with the erroneous statement in ShM option pricing formula. On where page 319, Hersh Shefrin wrote:

(HS-Claim) *"Suppose that investors agree on the risk-free process, and agree on the volatility of the risky asset, but disagree on the drift term for the risky asset. That is let investor 1 believe that the stock price S obeys the process $\frac{dS}{S} = \mu_1 dt + \sigma dZ$, where Z is a Winer process Let investor 2 believe that the stock price S obeys the process $\frac{dS}{S} = \mu_2 dt + \sigma dZ$ . How will options be priced in this framework? They will be priced according to Black-Scholes. Therefore, heterogeneity will not impact option prices, and will not give rise to volatility smiles."*

In general, HS-Claim is not true. To see that, suppose investor 1 takes the long position in the European option contract $\mathcal{C}$, with (1) price process $C(t) = f(S(t), t), t \geq 0$; (2) maturity time $T$; and (3) payoff function at maturity $f(S(T), T) = g(T)$, where $f(x, t), x > 0, t \geq 0$ is sufficiently

---

[15] See Barndorff-Nielsen and Shephard (2001a, b), Barndorff-Nielsen and Stelzer (2007), (2013), Muhle-Karbe, Pfaffel and Stelzer (2012),

smooth. Suppose investor 2 takes the short position in $C$. We make the following assumption $(AS1)$: *Suppose that investors 1 and 2 trade the asset according to their views and they know the views of each other*. If $(AS1)$ is not true what is the point of having different views on $S$, if the trading remains unchanged regarding the views of investors 1 and 2? Secondly, according ShM, there are only two investors in the market, so to believe that they do not see the history of trades of each other and blindly enter the option contract $C$, seems quite unrealistic. Assuming $(AS1)$, the dynamics of the long position in $C$ [16]is given by the Itô formula:

$$dC(t) = df(S(t), t) =$$

$$= \left( \frac{\partial f(S(t),t)}{\partial t} + \frac{\partial f(S(t),t)}{\partial x} \mu_1 S(t) + \frac{1}{2} \frac{\partial^2 f(S(t),t)}{\partial x^2} \sigma^2 S(t)^2 \right) dt + \frac{\partial f(S(t),t)}{\partial x} \sigma S(t) dZ(t).$$

Investor 2, forms a self-financing strategy $C(t) = f(S(t),t) = a(t)S(t) + b(t)\beta(t), t \geq 0$, where $\beta(t) = \beta(0)e^{rt}, t \geq 0$, is the riskless asset with a riskless rate $r$. Then the dynamics of the replicating portfolio is given by

$$dC(t) = df(S(t), t) = a(t)dS(t) + b(t)d\beta(t) =$$

$$= \left( a(t)\mu_2 S(t) + b(t)r\beta(t) \right) dt + a(t)\sigma S(t) dZ(t).$$

Equating the expressions for $dC(t)$ leads to $a(t) = \frac{\partial f(S(t),t)}{\partial x}$ and

$$\frac{\partial f(S(t),t)}{\partial t} + \frac{\partial f(S(t),t)}{\partial x} \mu_1 S(t) + \frac{1}{2} \frac{\partial^2 f(S(t),t)}{\partial x^2} \sigma^2 S(t)^2 = a(t)\mu_2 S(t) + r \left( f(S(t),t) - a(t)S(t) \right).$$

Setting $S(t) = x$, leads to the partial differential equation (PDE):

$$\frac{\partial f(x,t)}{\partial t} + \frac{\partial f(x,t)}{\partial x}(\mu_1 - \mu_2 + r)x - rf(x,t) + \frac{1}{2} \frac{\partial^2 f(x,t)}{\partial x^2} \sigma^2 x^2 = 0,$$

---

[16] We now follow the classical derivation of the Black-Scholes formula, see Black and Scholes (1973), and Duffie (2001), Section 5F.

Which is the Black-Scholes formula in the homogeneous case $\mu_1 = \mu_2$. In general, the above PDE will lead to option pricing with arbitrage opportunities.

We now start adjusting the HS-Claim to make it consistent with RDAPT. As with Shefrin (2005), Chapter 21, suppose there are two traders $\beth^{(j)}, j = 1,2$ who view and trade one risky asset (stock) $\mathcal{S}$ [17]. $\beth^{(j)}, j = 1,2$ trade $\mathcal{S}$ differently, because they have different views (estimates of the stock dynamics) and potentially having different trading skills. Both assume that $\mathcal{S}$-price dynamics is given by a geometric Brownian motion (GBM) but their opinions on the coefficients of the GBM differ: for $\beth^{(j)}$, the price dynamics of $\mathcal{S}$ is given by

$$S^{(j)}(t) = x^{(0)} \exp\left(\left(\mu^{(j)} - \frac{\sigma^2}{2}\right)t + \sigma B(t)\right), t \geq 0, \ x^{(0)} > 0, \mu^{(2)} > \mu^{(1)} > 0, \ \sigma > 0, \quad (25)$$

where $B(t), t \geq 0$, is a Brownian Motion generating a stochastic basis $(\Omega, \mathcal{F}, \mathbb{F} = (\mathcal{F}_t, t \geq 0), \mathbb{P})$.

Without the introduction of transaction costs eliminating in (25) the arbitrage opportunities, as we have shown already, the market model (25) is useless. Now our first task is to extend Black (1972)'s model on markets with no riskless asset. The extension consists of removing Black's assumption that asset volatilities are different. This will require the introduction of transaction costs on the velocity of hedging portfolio. We shall derive: (25) the general form of those transaction costs, and (2) the interest rate determining the discount factor the representative agent should be using[18].

---

[17] Such a difference in trading $\mathcal{S}$ could be due to (25) $\beth^{(1)}$'s and $\beth^{(2)}$'s different statistical estimations of the model parameters; $(ii)$ $\beth^{(1)}$'s and $\beth^{(2)}$'s are choosing different probability weighting functions (see Prelec (1998)) when they temper their views on $\mathcal{S}$-return diostribution; $(iii)$ $\beth^{(1)}$ and $\beth^{(2)}$ when trade $\mathfrak{W}$ they exhibit different level of trading performance.

### 3.1. *Market with Traders Having Heterogeneous Views on the Underlying Asset in the Presence of Arb- Costs; Determining of the Representative's Agent Riskless Rate*

Suppose (25) holds. The (hypothetical) representative agent (designated as $\aleph$ ) sees that market with price processes (25) allow for arbitrage opportunities. Realizing that, $\aleph's$ tasks are two: $(T1)$ Determine the general structure of arb-costs which will eliminate the gains from potential arbitrage opportunities and $(T2)$ determine the riskless rate $r^*$ in the market (25) with transaction costs given in $(T1)$.

Starting with $(T1)$, $\aleph$, knowing the price dynamics (25) for both traders $\beth^{(j)}, j = 1,2$ , decides to consider a hypothetical perpetual derivative contract, $\mathcal{G}$ , with price process $G(t) = g\left(S^{(1)}(t), S^{(2)}(t)\right), t \geq 0$, where $g: (0, \infty)^2 \to (0, \infty)$ has continuous second derivatives. As a representative agent, $\aleph$ takes simultaneously the long and the short position in $\mathcal{G}$. The long-position dynamics is determined by the Itô formula

---

[18] The formula (21.9) in Shefrin (2005) page 306 for the risk-free interest rate and formula (32) in Beninga and Mayshar (2000), Section V.a. for the representative agent 's discount factors are incorrect from the viewpoint of RDAPT, as they could imply arbitrage opportunities. One might argue that this is not of concern to behavioralists, but we argue that no-arbitrage option pricing should be "must" to traders. Suppose a trader is using a ShM-behavioral type option pricing in practice. Then, in real markets, there will be at least one" rational" trader who will explore the arb-opportunity, ShM-behavioral option pricing is generating.

$$dG(t) = dg\left(S^{(1)}(t), S^{(2)}(t)\right) =$$

$$= \begin{pmatrix} \dfrac{\partial g\left(S^{(1)}(t), S^{(2)}(t)\right)}{\partial x^{(1)}} \mu^{(1)} S^{(1)}(t) + \dfrac{\partial g\left(S^{(1)}(t), S^{(2)}(t)\right)}{\partial x^{(2)}} \mu^{(2)} S^{(2)}(t) + \\[4mm] + \dfrac{1}{2} \dfrac{\partial^2 g\left(S^{(1)}(t), S^{(2)}(t)\right)}{\partial x^{(1)^2}} \sigma^2 S^{(1)}(t)^2 + \dfrac{1}{2} \dfrac{\partial^2 g\left(S^{(1)}(t), S^{(2)}(t)\right)}{\partial x^{(2)^2}} \sigma^2 S^{(2)}(t)^2 + \\[4mm] + \dfrac{\partial^2 g\left(S^{(1)}(t), S^{(2)}(t)\right)}{\partial x^{(1)} \partial x^{(2)}} \sigma^2 S^{(1)}(t) S^{(2)}(t) \ ) \end{pmatrix} dt$$

$$+ \left( \dfrac{\partial g\left(S^{(1)}(t), S^{(2)}(t)\right)}{\partial x^{(1)}} \sigma S^{(1)}(t) + \dfrac{\partial g\left(S^{(1)}(t), S^{(2)}(t)\right)}{\partial x^{(2)}} \sigma S^{(2)}(t) \right) dB(t).$$

Knowing that the short position in $\mathcal{G}$ can employ arbitrage self-financing trading strategies, $\aleph$ decides to introduce arb-costs on the velocity of the hedge portfolio

$$P\left(S^{(1)}(t), S^{(2)}(t)\right) := a^{(1)}(t) S^{(1)}(t) + a^{(2)}(t) S^{(2)}(t) \equiv g\left(S^{(1)}(t), S^{(2)}(t)\right), t \geq 0,$$

that is, in the delta- position of $P\left(S^{(1)}(t), S^{(2)}(t)\right), t \geq 0$. More precisely, $\aleph$ defines the structure of arb-costs in the $P$-dynamics as follows:

$$dP\left(S^{(1)}(t), S^{(2)}(t)\right) = dg\left(S^{(1)}(t), S^{(2)}(t)\right) =$$

$$= \left( \lambda^{(1)} a^{(1)}(t) - \rho^{(1)} \dfrac{\partial g\left(S^{(1)}(t), S^{(2)}(t)\right)}{\partial x^{(1)}} \right) dS^{(1)}(t) +$$

$$+ \left( \lambda^{(2)} a^{(2)}(t) - \rho^{(2)} \dfrac{\partial g\left(S^{(1)}(t), S^{(2)}(t)\right)}{\partial x^{(1)}} \right) dS^{(2)}(t), \qquad (26)$$

where the transaction rates $\lambda^{(j)} > 0, \rho^{(j)} > 0, j = 1,2$ will be determined by the corresponding Black-Scholes type PDE guaranteeing a fair-price process $G(t) = g\left(S^{(1)}(t), S^{(2)}(t)\right), t \geq 0$ for $\mathcal{G}$. Thus, equating the terms with $dg\left(S^{(1)}(t), S^{(2)}(t)\right)$ leads to

$$a^{(1)}(t)S^{(1)}(t) = \frac{1}{\lambda^{(1)}\sigma^{(1)} - \lambda^{(2)}\sigma^{(2)}} \times$$

$$\times \left( \begin{array}{c} \dfrac{\partial g\left(S^{(1)}(t), S^{(2)}(t)\right)}{\partial x^{(1)}}\sigma S^{(1)}(t) + \dfrac{\partial g\left(S^{(1)}(t), S^{(2)}(t)\right)}{\partial x^{(2)}}\sigma S^{(2)}(t) + \\ + \rho^{(1)}\dfrac{\partial g\left(S^{(1)}(t), S^{(2)}(t)\right)}{\partial x^{(1)}}S^{(1)}(t)\sigma + \rho^{(2)}\dfrac{\partial g\left(S^{(1)}(t), S^{(2)}(t)\right)}{\partial x^{(2)}}S^{(2)}(t)\sigma - \\ - \lambda^{(2)} g\left(S^{(1)}(t), S^{(2)}(t)\right)\sigma \end{array} \right).$$

Applying $g\left(S^{(1)}(t), S^{(2)}(t)\right) = a^{(1)}(t)S^{(1)}(t) + a^{(2)}(t)S^{(2)}(t)$ results in

$$\frac{\mu^{(2)} - \mu^{(1)}}{\lambda^{(1)} - \lambda^{(2)}}\lambda^{(2)}\left(1 + \rho^{(1)}\right)S^{(1)}(t)\frac{\partial g\left(S^{(1)}(t), S^{(2)}(t)\right)}{\partial x^{(1)}} +$$

$$+ \frac{\mu^{(2)} - \lambda^{(1)}}{\lambda^{(1)} - \lambda^{(2)}}\lambda^{(1)}\left(1 + \rho^{(2)}\right)S^{(2)}(t)\frac{\partial g\left(S^{(1)}(t), S^{(2)}(t)\right)}{\partial x^{(2)}} +$$

$$- r^* g\left(S^{(1)}(t), S^{(2)}(t)\right) + \frac{1}{2}\frac{\partial^2 g\left(S^{(1)}(t), S^{(2)}(t)\right)}{\partial x^{(1)2}}\sigma^2 S^{(1)}(t)^2 = 0,$$

where

$$r^* = \frac{\mu^{(2)} - \mu^{(1)}}{\lambda^{(1)} - \lambda^{(2)}}\lambda^{(1)}\lambda^{(2)}. \tag{27}$$

Thus setting

$$\rho^{(j)} = 1 - \lambda^{(j)}, j = 1,2, \tag{28}$$

leads to the following Black-Scholes PDE for $g(x^{(1)}, x^{(2)}), x^{(1)} > 0, x^{(2)} > 0$,

$$r^* \frac{\partial g(x^{(1)}, x^{(2)})}{\partial x^{(1)}} x^{(1)} + r^* \frac{\partial g(x^{(1)}, x^{(2)})}{\partial x^{(1)}} x^{(2)} - r^* g(x^{(1)}, x^{(2)}) +$$

$$+ \frac{1}{2} \frac{\partial^2 g(x^{(1)}, x^{(2)})}{\partial x^{(1)^2}} \sigma^2 x^{(1)^2} + \frac{1}{2} \frac{\partial^2 g(x^{(1)}, x^{(2)})}{\partial x^{(2)^2}} \sigma^2 x^{(1)^2} +$$

$$+ \frac{\partial^2 g(x^{(1)}, x^{(2)})}{\partial x^{(1)} \partial x^{(2)}} ) \sigma^2 x^{(1)} x^{(2)} = 0. \tag{29}$$

The representation of the riskless rate (27) can be viewed as a generalization of Black (1972)'s formula for the riskless rate in the case of markets with transaction costs and equal stock-volatilities.

Now, the next task ℵ has is to determine the parameters $\lambda^{(1)} > 0$ and $\lambda^{(2)} > 0$.

### 3.2. *Market with Traders Having Heterogeneous Views on the Underlying Asset in the Presence of Arb- transaction Costs; Determining Model Parameters*

To determine the parameters $\lambda^{(j)} > 0, j = 1,2$, ℵ considers the market with price processes given by (25) and a riskless bond price

$$\beta(t) = \beta(0) e^{r^* t}, t \geq 0, \tag{30}$$

where the riskless rate $r^*$ is given by (27). ℵ decides to enter simultaneously the long and the short positions in a hypothetical perpetual derivative contracts $\mathcal{H}^{(j)}, j = 1,2$, with price processes $H^{(j)}(t) = h^{(j)} \left( S^{(j)}(t), \beta(t) \right), t \geq 0$, where $h^{(j)}(x, y), x > 0, \beta(t)$ has continuous derivatives and $\frac{\partial^2 h^{(j)}(x,t)}{\partial x^2}, \frac{\partial h^{(j)}(x,y)}{\partial y}$. By the Itô formula,

$$dH^{(j)}(t) = dh^{(j)} \left( S^{(j)}(t), \beta(t) \right) =$$

$$= \left( \frac{\partial h^{(j)}\left(S^{(j)}(t), \beta(t)\right)}{\partial x} S^{(j)}(t)\mu^{(j)} + \frac{\partial h^{(j)}\left(S^{(j)}(t), \beta(t)\right)}{\partial y} r\beta(t) + \frac{\partial^2 h^{(j)}\left(S^{(j)}(t), \beta(t)\right)}{\partial x^2} \sigma^2 S^{(j)}(t)^2 \right) dt$$

$$+ \frac{\partial h^{(j)}\left(S^{(j)}(t), \beta(t)\right)}{\partial x} S^{(j)}(t)\sigma dB(t).$$

ℵ forms a replicating portfolio, $P^{(j)}(t) := a^{(j)}(t)S^{(j)}(t) + b^{(j)}(t)\beta(t) \equiv h^{(j)}\left(S^{(j)}(t), t\right), t \geq 0.$

Now ℵ defines transactions costs in the hedged portfolio as follows:

$$dP^{(j)}(t) = dh^{(j)}\left(S^{(j)}(t), \beta(t)\right) =$$

$$= \left( \lambda^{(j)} a^{(j)}(t) - \left(1 - \lambda^{(j)}\right) \frac{\partial h^{(j)}\left(S^{(j)}(t), \beta(t)\right)}{\partial x} \right) dS^{(j)}(t) +$$

$$+ \left( \lambda^{(bond)} b(t) - \left(1 - \lambda^{(bond)}\right) \frac{\partial h^{(j)}\left(S^{(j)}(t), \beta(t)\right)}{\partial y} \right) d\beta(t).$$

Equating the terms with $dh^{(j)}\left(S^{(j)}(t), t\right), t \geq 0$, leads to $a^{(j)}(t) = \frac{2 + \lambda^{(j)}}{\lambda^{(j)}} \frac{\partial h^{(j)}(S^{(j)}(t), t)}{\partial x}$

and

$$\left( \mu^{(j)} - (1 + 2\lambda^{(j)})\mu^{(j)} + \lambda^{(bond)} r^* \frac{2 + \lambda^{(j)}}{\lambda^{(j)}} \right) \frac{\partial h^{(j)}\left(S^{(j)}(t), \beta(t)\right)}{\partial x} S^{(j)}(t)$$

$$+ \left( \left(1 - \lambda^{(bond)}\right) r^* \right) \frac{\partial h^{(j)}\left(S^{(j)}(t), \beta(t)\right)}{\partial y} \beta(t) +$$

$$-\lambda^{(bond)} r^* h^{(j)}\left(S^{(j)}(t), \beta(t)\right) + \frac{1}{2} \frac{\partial^2 h^{(j)}\left(S^{(j)}(t), \beta(t)\right)}{\partial x^2} \sigma^2 S^{(j)}(t)^2 = 0.$$

Thus, $\lambda^{(bond)}$ should be set to 1. Furthermore, from $\mu^{(j)} - (1 + 2\lambda^{(j)})\mu^{(j)} + \lambda^{(bond)} r^* \frac{2 + \lambda^{(j)}}{\lambda^{(j)}} = r^*$, it follows that $\lambda^{(j)} = \sqrt{\frac{r^*}{\mu^{(j)}}}$ which together with (27), implies that

$$r^* = \left( \sqrt{\mu^{(2)}} + \sqrt{\mu^{(1)}} \right)^2, \quad \lambda^{(j)} = \frac{\sqrt{\mu^{(2)}} + \sqrt{\mu^{(1)}}}{\sqrt{\mu^{(j)}}}, j = 1,2. \tag{31}$$

Setting $S^{(j)}(t) = x$, we obtain the Black-Scholes type PDE for $h^{(j)}(x,y), x > 0, y > 0$

$$r^* \frac{\partial h^{(j)}(x,y)}{\partial x} x + r^* \frac{\partial h^{(j)}(x,y)}{\partial y} y - r^* h^{(j)}(x,y) + \frac{1}{2}\sigma^2 \frac{\partial^2 h^{(j)}(x,y)}{\partial x^2} x^2 = 0. \qquad (32)$$

Summarizing our findings, we formulate the following proposition.

PROPOSITION 2: *Suppose* $\beth^{(j)}, j = 1,2$ *trade a single risky asset* $\mathcal{S}$ *with different price dynamics given by (25). Then the no-arbitrage riskless* $r^*$ *is given by (31) and the bond dynamics* $\beta(t), t \geq 0,$ *is given by (30). Next, consider a perpetual derivative contract* $\mathfrak{W}$, *with price process* $\mathcal{W}(t) = W(S(t), t),$ *where* $S(t), t \geq 0,$ *is either* $S^{(1)}(t), t \geq 0$ *or* $S^{(2)}(t), t \geq 0,$ *and* $W(x,t), x > 0, t \geq 0$ *has continuous* $\frac{\partial W(x,t)}{\partial t}$ *and* $\frac{\partial^2 W(x,t)}{\partial x^2}$. *If* $\beth^{(j)}, j = 1,2$ *is taking the short position in* $\mathfrak{W}$, $\beth^{(j)}$'s *replicating hedge-portfolio* $P^{(j)}(t) := a^{(j)}(t)S^{(j)}(t) + b^{(j)}(t)\beta(t) \equiv W(S^{(j)}(t), t), t \geq 0,$ *is subject to arb-costs of the following type:*

$$dP^{(j)}(t) := \left( \lambda^{(j)} a^{(j)}(t) - \left(1 - \lambda^{(j)}\right) \frac{\partial W(S^{(j)}(t),t)}{\partial x} \right) dS^{(j)}(t) + b^{(j)}(t)d\beta(t), \qquad (33)$$

*where* $\lambda^{(j)}, j = 1,2$ *are given by (31). Then* $W(x,t), x > 0, t \geq 0,$ *satisfies the Black-Scholes type PDE:*

$$\frac{\partial W(x,t)}{\partial t} + r^* \frac{\partial W(x,t)}{\partial x} x - r^* W(x,t) + \frac{1}{2}\sigma^2 \frac{\partial^2 W(x,t)}{\partial x^2} x^2 = 0. \qquad (34)$$

The proof of Proposition 2 follows directly from (31) and (32).

Formula (34) adjusts HS-Clam to be consistent with RDAPT, by introducing special form of arb-costs. In the next two sections, we continue studying HS-Claim with different type of arb-costs.

## 4. Option Pricing with Heterogeneous Views on Binomial Lattice

Suppose the traders $\beth^{(S)}$ and $\beth^{(V)}$ observe each other's trading history. $\beth^{(S)}$ (resp. $\beth^{(V)}$) trades a risky asset (stock) $\mathfrak{W}$ on a binomial lattice with price dynamics $S_{k\Delta t}, k \in \mathcal{N}^{(0)} = \{0,,1,\dots,\}, S_0 > 0$ (resp. $V_{k\Delta t}, k \in \mathcal{N}^{(0)}, V_0 > 0$ ). The joint price process dynamics is given by the binomial tree[19]:

$$\begin{bmatrix} S_{(k+1)\Delta t} \\ V_{(k+1)\Delta t} \end{bmatrix} = \begin{cases} \begin{bmatrix} S_{(k+1)\Delta t;up} = S_{k\Delta t}\left(1 + \mu\Delta t + \sigma\sqrt{\Delta t}\right) \\ V_{(k+1)\Delta t;up} = V_{k\Delta t}\left(1 + m\Delta t + v\sqrt{\Delta t}\right)v\sqrt{\Delta t} \end{bmatrix} w.p.\frac{1}{2} \\ \begin{bmatrix} S_{(k+1)\Delta t;down} = S_{k\Delta t}\left(1 + \mu\Delta t - \sigma\sqrt{\Delta t}\right) \\ V_{(k+1)\Delta t;down} = V_{k\Delta t}\left(1 + m\Delta t - v\sqrt{\Delta t}\right) \end{bmatrix} w.p.\frac{1}{2} \end{cases} \qquad (35)$$

$k \in \mathcal{N}^{(0)}, \Delta t > 0, \mu \in \mathcal{R}, m \in \mathcal{R}, \sigma > 0, v > 0$. For every fixed $T > 0$, the bivariate binomial tree $(S_{k\Delta t}, V_{k\Delta t})_{k \in 0,\dots,N\Delta t}$ generates a bivariate polygon process with trajectories in the Prokhorov space $C([0,T]^2)$ which converges weakly to the following bivariate geometric Brownian motion $(S_t, V_t)_{t \in [0,T]}$[20] :

$$S_t = S_0 e^{\left(\mu - \frac{\sigma^2}{2}\right)t + \sigma B(t)}, V_t = V_0 e^{\left(m - \frac{v^2}{2}\right)t + v B(t)}, t \in [0,T], \qquad (36)$$

where $B(t), t \geq 0$, is a Brownian Motion generating a stochastic basis $(\Omega, \mathcal{F}, \mathbb{F} = (\mathcal{F}_t, t \geq 0), \mathbb{P})$.

To derive the state price dynamics of the representative investor $\aleph$, let us consider a perpetual European derivative contract, $\mathcal{G}$. $\mathcal{G}$ has price process $G_{k\Delta t} = G(S_{k\Delta t}, V_{k\Delta t}), k \in \mathcal{N}^{(0)}$.

---

[19] This binomial tree (11) was introduced in Kim at al (2016) (see also Jarrow and Rudd (2008)) as an extension of the classical CRR-model Cox, Ross and Rubinstein M. (1979). We use this more general binomial pricing tree, because we require the bivariate pricing tree to be driven by one risk factor, and with that requirement, CRR-model is not appropriate.

[20] The proof is similar to that in Davydov and Rotar (2008), Theorem 2, and thus is omitted.

We assume that $\aleph$ is observing historical trading activities of $\beth^{(S)}$ and $\beth^{(V)}$. $\aleph$ (as an representative agent) has taken simultaneously both, the long and the sort position in, $\mathcal{G}$. $\aleph$ trades $S_t$ (resp. $V_t$) as $\beth^{(S)}$ (resp. $\beth^{(V)}$) would do. $\aleph$ forms a self-financing strategy $(a_{k\Delta t}, b_{k\Delta t}), k \in \mathcal{N}$ generating a self-financing portfolio $P(t) = a(t)S(t) + b(t)V(t)$. Thus, $G(t^{(k)}) = g\left(S(t^{(k)}), V(t^{(k)})\right) = P(t^{(k)}) = a(t^{(k)})S(t^{(k)}) + b(t^{(k)})V(t^{(k)})$.

When $\aleph$ trades as $\beth^{(S)}\left(resp.\beth^{(V)}\right)$), at any time interval $[t^{(k+1)}, t^{(k+1)})$ the trade is subject to transaction cost: $\left(\frac{S(t^{(k+1)})}{S(t^{(k)})}\right)^{\rho(\mathcal{S})} = 1 + \rho(\mathcal{S})ln\frac{S(t^{(k+1)})}{S(t^{(k)})}$, $\left(resp.\left(\frac{V(t^{(k+1)})}{\nu(t^{(k)})}\right)^{\rho(\mathcal{V})} = 1 + \rho(\mathcal{V})ln\frac{\nu(t^{(k+1)})}{V(t^{(k)})}\right)$ where $\rho(\mathcal{S}) = \mathfrak{C}\frac{\mu}{\sigma}\left(resp.\rho(\mathcal{S}) = \mathfrak{C}\frac{m}{v}\right)$, and $\mathfrak{C}$ is an absolute constant, .

Next, $\aleph$ chooses $\left(a(t^{(k)}), b(t^{(k)})\right)$, so that $-g\left(S(t^{(k+1)}), V(t^{(k+1)})\right) + a(t^{(k)})S(t^{(k+1)})\left(\frac{S(t^{(k+1)})}{S(t^{(k)})}\right)^{\rho(\mathcal{S})} + b(t^{(k)})V(t^{(k+1)})\left(\frac{\nu(t^{(k+1)})}{\nu(t^{(k)})}\right)^{\rho(\mathcal{V})} = 0$. That is,

$$a(t^{(k)})S(t^{(k+1,up)})\left(\frac{S(t^{(k+1,up)})}{S(t^{(k)})}\right)^{\rho(\mathcal{S})} + b(t^{(k)})\left(\frac{V(t^{(k+1,up)})}{\nu(t^{(k)})}\right)^{\rho(\mathcal{V})} =$$

$$= g\left(S(t^{(k+1,up)}), V(t^{(k+1,up)})\right) \tag{37}$$

and $a(t^{(k)})S(t^{(k+1,down)})\left(\frac{S(t^{(k+1,down)})}{S(t^{(k)})}\right)^{\rho(\mathcal{S})} + b(t^{(k)})V(t^{(k+1,down)})\left(\frac{V(t^{(k+1,down)})}{\nu(t^{(k)})}\right)^{\rho(\mathcal{V})} =$ $g\left(S(t^{(k+1,down)}), V(t^{(k+1,down)})\right)$. Thus, solving the two equations for $a(t^{(k)})$ and $b(t^{(k)})$, and evaluating $G(t^{(k)}) = a(t^{(k)})S(t^{(k)}) + b(t^{(k)})V(t^{(k)})$, $\aleph$ obtains the binomial option price dynamics:

$$g\left(S(t^{(k)}), V(t^{(k)})\right) = Q^{(\Delta t)}g(t^{(k+1,up)}) + (1 - Q^{(\Delta t)})g(t^{(k+1,down)}) \tag{38}$$

where the risk-neutral probabilities ($\aleph$'s state-price probabilities) are

$$Q^{(\Delta t)} = \frac{1}{2} - \frac{\mu\left(1+\mathfrak{C}\frac{\mu}{\sigma}\right)\left(1+\mathfrak{C}\frac{\sigma}{2}\right) - m\left(1+\mathfrak{C}\frac{m}{v}\right)\left(1+\mathfrak{C}\frac{v}{2}\right)}{2\left(\sigma - v + \mathfrak{C}(\mu - m)\right)}\sqrt{\Delta t} \tag{39}$$

and $1 - Q^{(\Delta t)}$. Note that even if (1) $\sigma = v$ (which is an arbitrage pricing model, if $\beth^{(S)}$ and $\beth^{(V)}$ trades without arb-costs), as soon as (2) $\mu \neq m$, risk neutral probabilities exist, and (3) the transaction costs have offset the arbitrage gains. From Kim at al. (2016), Section 3.2, and Black (1972), $Q^{(\Delta t)}$ should have the representation

$$Q^{(\Delta t)} = \frac{1}{2} - \frac{1}{2}\frac{\mu^{(*)} - r^{(*)}}{\sigma^{(*)}}\sqrt{\Delta t} = \frac{1}{2} - \frac{1}{2}\frac{m^{(*)} - r^{(*)}}{v^{(*)}}\sqrt{\Delta t}, \tag{40}$$

where

$$r^{(*)} := \frac{\mu^{(*)}v^{(*)} - m^{(*)}\sigma^{(*)}}{v^{(*)} - \sigma^{(*)}}, \ \mu^{(*)} := \mu\left(1 + \mathfrak{C}\frac{\mu}{\sigma}\right)\left(1 + \mathfrak{C}\frac{\sigma}{2}\right),$$

$$m^{(*)} := m\left(1 + \mathfrak{C}\frac{m}{v}\right)\left(1 + \mathfrak{C}\frac{v}{2}\right), \ \sigma^{(*)} := \sigma + \mathfrak{C}\mu, v^{(*)} = v + \mathfrak{C}m. \tag{41}$$

In (41), $r^{(*)}$ is $\aleph$'s risk neutral rate, and $\mu^{(*)}, m^{(*)}, \sigma^{(*)}$ and $v^{(*)}$ are the adjusted (for arb-cost) drift and volatility parameters, satisfying (39) and (40). Now, the price processes $(S_t, V_t)_{t \in [0,T]}$, as seen by $\aleph$ in the risk-neutral world $(\Omega, \mathcal{F}, \mathbb{F} = (\mathcal{F}_t, t \geq 0), \mathbb{Q})$, $\mathbb{Q} \sim \mathbb{P}$ have the price dynamics:

$$S_t = S_0 e^{\left(r^{(*)} - \frac{\sigma^{(*)2}}{2}\right)t + \sigma^{(*)}B^{(*)}(t)}, V_t = V_0 e^{\left(r^{(*)} - \frac{v^{(*)2}}{2}\right)t + v^{(*)}B^{(*)}(t)}, t \in [0, T], \tag{42}$$

$B^{(*)}(t), t \geq 0$, is a Brownian motion on $\mathbb{Q}$, and an arithmetic Brownian motion on $\mathbb{P}$ with

$B^{(*)}(t) = B(t) + \theta^{(*)}t$. The parameter $\theta^{(*)} = \frac{\mu^{(*)} - r^{(*)}}{\sigma^{(*)}} = \frac{m^{(*)} - r^{(*)}}{v^{(*)}}$ is the market price of risk in $\aleph's$ market model with arb-costs.

Note that in the case of trading with without arb-costs, $Q^{(\Delta t; no\ arb-cost)}$

$:= \frac{1}{2} - \frac{\mu - m}{2(\sigma - v)} \sqrt{\Delta t}$. Thus, $Q^{(\Delta t; no\ arb-cost)}$ is the risk-neural probability in the Black(1972) model:

$Q^{(\Delta t; no\ trans\ cost)} := \frac{1}{2} - \frac{1}{2} \frac{\mu - r}{\sigma} \sqrt{\Delta t}$, where $\frac{\mu - r}{\sigma} = \frac{m - r}{v} = \frac{\mu - m}{\sigma - v}$ and $r = \frac{\mu v - m\sigma}{v - \sigma}$. Thus, $\aleph's$ model can be viewed as an extension of the Black (1972) model when special type of transaction costs is introduced.

**5. Black-Scholes Formula for Markets with Arb- Transaction Costs in the Presence of Transaction Costs on the Delta-and Gamma-positions**

In this section, we extend our binomial tree model with arb- costs in Section 3, to market with continuous time price process. $\beth^{(S)}$ and $\beth^{(V)}$ view and trade one share of the price process $\mathfrak{W}$ as different diffusions price processes:

$$dS_t = \mu_t S_t dt + \sigma_t S_t dB_t, S_0 > 0, \ dV_t = m_t S_t dt + v_t S_t dB_t, S_0 > 0, \tag{43}$$

where

$(Sec1a): B_t, t \geq 0$, is a standard Brownian motion generating stochastic basis $(\Omega, \mathcal{F}, \mathbb{F} = (\mathcal{F}_t, t \geq 0), \mathbb{P})$;

$(Sec1b): \mu_t = \mu(t, S_t), t \geq 0$ and $\sigma_t = \sigma(t, S_t), t \geq 0$, are $\mathbb{F}$- adapted drift and diffusion coefficients satisfying the regularity conditions guaranteeing existence and uniqueness of the strong solution of $(43)$[21];

---

[21] See for example, Duffie (2001) Chapter 6, Shreve (2004), Chapter 4, Shiryev (1999) Chapter3

$(Sec2)$: a risk-free asset (bond) $\mathfrak{B}$, with price process $\beta_t, t \geq 0$, following continuous diffusion:

$$d\beta_t = r_t \beta_t dt, t \geq 0, \tag{44}$$

where $r_t = r(t, S_t) > 0, t \geq 0$ is $\mathbb{F}$- adapted, risk-free rate satisfying the regularity condition:

$\sup\left\{r_t + \frac{1}{r_t}; t \geq 0\right\} < \infty, \mathbb{P}$-a.s.

Consider a portfolio, $\mathcal{X}$, with price dynamics

$$X_t = \mathbb{X}(t, S_t) = a_t S_t + b_t \beta_t, t \geq 0, \tag{45}$$

where

$(P1)$: $\mathbb{X}(t, x), t \geq 0, x > 0$, has continuous derivatives $\frac{\partial \mathbb{X}(t,x)}{\partial t}$ and $\frac{\partial^2 \mathbb{X}(t,x)}{\partial x^2}, t \geq 0, x > 0$;

$(P1)$: the trading strategy $(a_t = a(t, S_t), b_t = b(t, S_t); t \geq 0)$ is $\mathbb{F}$- adapted, and subject to *arb-cost*, $\Theta(t, S_t) = \left(\Delta(t, S_t), \Gamma(t, S_t), \Psi(t, S_t), \mathcal{C}(t, S_t)\right), t \geq 0$, a $\mathbb{F}$- adapted quadruplet, where

$(P1a)$ $\Delta(t, S_t) \in (0,1), t \geq 0$, is the *delta*-cost;

$(P1b)$ $\Gamma(t, S_t) > 0, t \geq 0$, is the *gamma*-cost;

$(P1c)$ $\Psi(t, S_t) \in (0,1), t \geq 0$, is the cost associated with trading the bond;

$(P1d)$ $\mathcal{C}(t, S_t) = \hbar(t, S_t) - \kappa(t, S_t), t \geq 0$, is the consumption cost $\hbar(t, S_t), t \geq 0$ minus the cost opportunities $\kappa(t, S_t) = \frac{\Delta(t,S_t)}{1-\Delta(t,S_t)}\left(1 - \Psi(t, S_t)\right)r_t \mathbb{X}(t, S_t)$ associated with eliminating the arbitrage opportunities;

$(P2)$: the $\mathcal{X}$-dynamics is given by

$$dX_t = d\mathbb{X}(t, S_t) = \left( a_t \left( 1 - \Delta\left(t, S_t\right) \right) - \frac{\sigma_t^2}{2} \Gamma\left(t, S_t\right) \frac{\partial^2 \mathbb{Y}(t, S_t)}{\partial x^2} S_t \right) dS_t +$$

$$+ b_t \left( 1 - \Psi\left(t, S_t\right) \right) d\beta_t - \mathcal{C}\left(t, S_t\right) dt, t \geq 0; \tag{46}$$

$(P3)$: the discount rate $\lambda_t, t \geq 0$, associated with $\mathcal{X}$-dynamics is given by

$$\lambda_t = \lambda(t, S_t) = r_t \frac{1 - \Psi\left(t, x\right)}{1 - \Delta\left(t, x\right)} > 0, t \geq 0; \tag{47}$$

$(P5)$: the augmented volatility $\rho(t, x)$, associated with $\mathcal{X}$-dynamics is given by

$$\rho(t, x) = \sqrt{1 + \lambda(t, S_t) \Gamma\left(t, x\right)} \, \sigma(t, x) > 0, t \geq 0, x > 0. \tag{48}$$

The trading strategy $(a_t, b_t), t \geq 0$, satisfying (45) - (48), is an *arbitrage*, if the $\mathcal{X}$-dynamics generated by $(a_t, b_t), t \geq 0$ satisfies: $\mathbb{P}(X_0 \leq 0, X_T \geq 0) = 1$, and $\mathbb{P}(X_T > 0) > 0$ for some $T > 0$.

Consider a new security in the market $(\mathfrak{S}, \mathfrak{B})$, an European derivative $\mathcal{Y}$, with price process $Y_t = \mathbb{Y}(t, S_t)$, where $\mathbb{Y}(t, x), t \geq 0, x > 0$ has continuous derivatives $\frac{\partial \mathbb{Y}(t, x)}{\partial t}$ and $\frac{\partial^2 \mathbb{Y}(t, x)}{\partial x^2}, t \geq 0, x > 0$. The payoff of $\mathcal{Y}$ at the expiration date $T > 0$ is $Y_T = \mathbb{Y}(T, S_T) = g(S_T)$, for some continuous $g(x), x \geq 0$.

By the Itô formula:

$$d\mathbb{Y}(t, S_t) = \left\{ \frac{\partial \mathbb{Y}(t, S_t)}{\partial t} + \frac{\partial \mathbb{Y}(t, S_t)}{\partial x} \mu_t S_t + \frac{1}{2} \frac{\partial^2 \mathbb{Y}(t, S_t)}{\partial x^2} \sigma_t^2 S_t^2 \right\} dt + \frac{\partial \mathbb{Y}(t, S_t)}{\partial x} \sigma_t S_t dB_t.$$

Consider a replicating portfolio $Y_t = \mathbb{Y}(t, S_t) = a_t S_t + b_t \beta_t, t \geq 0,$ satisfying

$$dY_t = d\mathbb{Y}(t, S_t) = \left( a_t \left( 1 - \Delta\left(t, S_t\right) \right) - \frac{\sigma_t^2}{2} \Gamma\left(t, S_t\right) S_t \frac{\partial^2 \mathbb{Y}(t, S_t)}{\partial x^2} \right) dS_t +$$

$$+b_t\big(1-\Psi\,(t,S_t)\big)d\beta_t-\left(\frac{\Delta\,(t,S_t)}{1-\Delta\,(t,S_t)}\big(1-\Psi\,(t,S_t)\big)r_t\mathbb{Y}(t,S_t)\,-\hbar\,(t,S_t)\right)dt$$

We have two equations for the $\mathcal{Y}$- dynamics, which together $\mathbb{Y}(t,S_t)=a_tS_t+b_t\beta_t$ give expressions for the trading strategies

$$a_t=\frac{1}{1-\Delta\,(t,S_t)}\left(\frac{\partial\mathbb{Y}(t,S_t)}{\partial x}+\frac{\sigma_t^2}{2}\Gamma\,(t,S_t)S_t\frac{\partial^2\mathbb{Y}(t,S_t)}{\partial x^2}\right),$$

and

$$b_t=\frac{1}{\beta_t}\left(\mathbb{Y}(t,S_t)-\frac{1}{1-\Delta\,(t,S_t)}\left(\frac{\partial\mathbb{Y}(t,S_t)}{\partial x}S_t+\frac{\sigma_t^2}{2}\Gamma\,(t,S_t)S_t^2\frac{\partial^2\mathbb{Y}(t,S_t)}{\partial x^2}\right)\right).$$

Substituting the expressions for $a_t$ and $b_t$ in the equation the drift of $\mathbb{Y}(t,S_t),t\geq0$, leads to the following proposition:

PROPOSITION 3. *Consider a market with securities* $(\mathfrak{S},\mathfrak{B},\mathcal{Y})$ *and suppose conditions (43)-(48)) hold. Then* $\mathbb{Y}(t,x),t\geq0,x>0$, *satisfies the partial differential equation: for all* $t\in[0,t),x>0$,

$$\frac{\partial\mathbb{Y}(t,x)}{\partial t}+\lambda(t,S_t)\frac{\partial\mathbb{Y}(t,x)}{\partial x}x-\lambda(t,S_t)\mathbb{Y}(t,x)+\frac{1}{2}\frac{\partial^2\mathbb{Y}(t,x)}{\partial x^2}\rho(t,x)^2x^2+\hbar\,(t,x)=0 \qquad (49)$$

*with boundary condition* $\mathbb{Y}(T,x)=g(x),x>0$.

The solution of (49) is given by the Feynman-Kac formula[22]:

$$\mathbb{Y}(t,x)=\mathbb{E}_{x,t}\left[\int_t^T\varphi_{t.s}\,\hbar\,(s,Z_s)ds+\varphi_{t.T}g(Z_T)\right], \qquad (50)$$

---

[22] See, for example, Duffie (2001), Section E.6.

where

$(FC1): Z_s, s \geq t, Z_t = x$, is a continuous diffusion given by

$$dZ_s = \lambda(s, Z_s)Z_s ds + \rho(s, Z_s)Z_s dW_s, s \geq t, \tag{51}$$

where $W_t, t \geq 0$ is a Brownian motion of a stochastic basis $(\Omega, \mathcal{F}, \mathbb{F} = (\mathcal{F}_t, t \geq 0), \mathbb{Q})$

$(FC2): \varphi_{t.s} = \exp\{-\int_t^s \lambda(u, Z_u) du\}, 0 \leq t < s$, is the family of discount factors;

$(FC3)$ $\mathbb{E}_{x,t}$ stands for the expected value indicating that the process $Z_s, s \geq t$ starts at $Z_t = x$.

## 6. Conclusions

In this paper, we correct statements made in Shefrin (2005) on behavioral option pricing. We pointed out that option pricing formulas in Shefrin (2005) and Benninga and Mayshar (2000) are with flaws from the viewpoint of rational dynamic asset pricing theory. Using those incorrect asset pricing formulas could lead to serious losses of the traders, as in the market there will be "rational" traders who will explore the arbitrage opportunities those behavioral option prices generates. In order to correct Shefrin's and Benninga and Mayshar's (2000)'s behavioral approaches to option pricing, we introduce several types of arb-costs on the velocity of hedged portfolio, so that the generated arbitrage gains be eliminated by those arb-costs. In case the market model without arbitrages, the introduction of the transaction costs on the velocity of hedged portfolio increases slightly the volatility of the hedge and indeed worsen the Sharpe ratio of the trader. Since it will be impossible to introduce transaction costs only when the trades is an arbitrage (as in the behavioral option pricing), we argue that a universal transaction costs on the velocity of trades should be introduced in real markets.